\documentclass[a4paper,12pt]{article}
\usepackage[top=1.25in, bottom=1.25in, left=1.05in, right=1.05in]{geometry}
\usepackage[utf8]{inputenc}
\usepackage[T1]{fontenc}
\usepackage[english]{babel}

\usepackage{natbib}
\usepackage[section]{placeins}
\usepackage{setspace, comment}
\usepackage{enumitem}
\usepackage{pdfpages, subfigure, tasks, graphicx}
\usepackage{pdflscape}
\usepackage{amscd,amsmath,amssymb,amsfonts,amsthm,bbm,bm,latexsym,mathrsfs,mathtools}
\usepackage[subnum]{cases}	% multi-case equations with separate number for each
\usepackage{dsfont}
\makeatletter
\newcommand*{\distas}[1]{\mathbin{\overset{#1}{\kern\z@\sim}}}	
\makeatother
\newcommand*\abs[1]{\left|#1\right|}		% absolute value
\allowdisplaybreaks

\usepackage[colorlinks=true,urlcolor=blue,linkcolor=blue,citecolor=blue]{hyperref}

\usepackage{color,graphicx}
\usepackage[width=0.9\textwidth, font={footnotesize}]{caption}

\usepackage{rotating,lscape,pdflscape}
\usepackage{booktabs,longtable,float,array,multirow,colortbl,adjustbox,threeparttable}

\usepackage[colorinlistoftodos,color=orange!60]{todonotes}
\usepackage{regexpatch}
\makeatletter
\xpatchcmd{\@todo}{\setkeys{todonotes}{#1}}{\setkeys{todonotes}{inline,#1}}{}{}
\makeatother

\newcommand\fnote[1]{\captionsetup{font=small}\caption*{#1}}

\makeatletter
\def \by{\mathbf{Y}}
\def \by{\mathbf{y}}

\def \bx{\mathbf{x}}

\def \bp{\mathbf{p}}

\def \bs{\mathbf{s}}

\def \bA{\mathbf{A}}
\def \bQ{\mathbf{Q}}
\def \bS{\mathbf{S}}
\def \bW{\mathbf{W}}
\def \bZ{\mathbf{Z}}

\def \bbeta{\boldsymbol{\beta}}
\def \bsigma{\boldsymbol{\sigma}}
\def \bomega{\boldsymbol{\omega}}

\def \bmu{\boldsymbol{\mu}}
\def \brho{\boldsymbol{\rho}}
\def \bxi{\boldsymbol{\xi}}
\def \bXi{\boldsymbol{\Xi}}
\def \bOmega{\boldsymbol{\Omega}}

\def \R {\mathds{R}}

\makeatother

\title{\vspace{-60pt} \textbf{A Bayesian Markov-switching SAR model for time-varying cross-price spillovers}
}

\author{
Matteo Iacopini\thanks{Queen Mary University of London, London, United Kingdom. \color{blue}\texttt{m.iacopini@qmul.ac.uk}}
\and
Christian Glocker\thanks{Austrian Institute of Economic Research, Vienna, Austria. \color{blue}\texttt{christian.glocker@wifo.ac.at}}
\and
Tamás Krisztin\thanks{International Institute for Applied Systems Analysis, Laxenburg, Austria. \color{blue}\texttt{krisztin@iiasa.ac.at}}
\and
Philipp Piribauer\thanks{Austrian Institute of Economic Research, Vienna, Austria. \color{blue}\texttt{philipp.piribauer@wifo.ac.at}}
}

\date{\today}

\begin{document}

\maketitle

\begin{abstract}
The spatial autoregressive (SAR) model is extended by introducing a Markov switching dynamics for the weight matrix and spatial autoregressive parameter.
The framework enables the identification of regime-specific connectivity patterns and strengths and the study of the spatiotemporal propagation of shocks in a system with a time-varying spatial multiplier matrix.
The proposed model is applied to disaggregated CPI data from 15 EU countries to examine cross-price dependencies. The analysis identifies distinct connectivity structures and spatial weights across the states, which capture shifts in consumer behaviour, with marked cross-country differences in the spillover from one price category to another.
% Words: 98

% Old version:
% Traditional models for spatially referenced data link the contemporaneous connectivity patterns in the data to a known, time-invariant spatial weight matrix. However, using an arbitrarily chosen notion of distance may hamper the results if the selected distance is not meaningful in capturing the connectedness of the data. Moreover, repeated observations of spatially referenced data over time highlight the importance of accounting for time changes in spatial dependence. 
% A new Markov-switching spatial autoregressive model is proposed by allowing the weight matrix to be estimated from the data and discretely vary over time, together with the strength parameter. The framework permits the identification of regime-specific connectivity patterns and strength, along with investigating the spatiotemporal propagation of shocks in the system due to the time-varying spatial multiplier matrix. The method is applied to sector-level price data from different countries in the EU ........

\vskip 8pt
\noindent \textbf{Keywords:} Bayesian inference; Markov switching; Random spatial weights; Spatial autoregressive model; Time-varying network.

\vskip 8pt
\noindent \textbf{JEL-Codes:} C23; C33; C34; E31 

\end{abstract}

\clearpage
\onehalfspacing

\section{Introduction}   \label{sec:introduction}

A cornerstone model within the spatial econometrics literature is the spatial autoregressive (SAR) model, which exploits a weight matrix to introduce spatial dependence among the observations. Empirical practise typically relies on the fundamental assumption of an exogenously given spatial weight matrix, whose entries are known and based on a neighbourhood concept between any two (spatial) observations, such as geographic proximity \citep{lesage2009introduction}. For a given definition of distance, the associated spatial matrix can be interpreted as representing the connectivity pattern (or network) according to that criterion.

Spatial autoregressive (SAR) models have been applied in statistics and econometrics to investigate diverse problems, including the measurement of sovereign risk spillovers \citep{debarsy2018measuring}, regional economic dynamics \citep{basile2014modeling}, and firm-level productivity spillovers \citep{baltagi2016firm}.
% Grouped SAR panel model \citep{huang2023grouped}
%
The importance of accounting for spatial dependence goes beyond linear conditional mean models. In particular, \cite{tacspinar2021bayesian} recently defined a spatial autoregressive stochastic volatility model to investigate volatility clustering in spatial data, while \cite{dougan2018bayesian} introduced spatial dependence in a sample selection model.
In addition, \cite{badinger2010horizontal} applied an extension of the SAR model to disentangle the horizontal (demand for final products) and vertical (demand for intermediate products) interdependence in multinational enterprise activity.

Several attempts have been made to relax the crucial assumption in the baseline SAR model of an exogenously fixed spatial weight matrix.
For instance, \cite{qu2015estimating} proposed a framework where the spatial weight is assumed to depend on an observable random variable and inferred by means of a generalised method of moments estimator, whereas \cite{qu2021estimation} utilised a two-stage estimation approach. Instead, \cite{krisztin2022bayesian} design a fully Bayesian approach to learn the unknown spatial weight matrix, thus allowing for uncertainty quantification on the dependence structure.
An alternative statistical approach consists of estimating SAR models using different types of spatial weight matrices and then applying model averaging procedures to address model uncertainty regarding selecting the best matrix \citep{zhang2018spatial,debarsy2022bayesian}.

Over the last decades, the use of weight matrices based solely on the geographical neighbourhood has been criticised \citep{corrado2012economics} due to the application of spatial regression models to contexts where the geographical location of the observations is not fundamental in driving capturing their connectivity structures, such as in trade, migration, and social networks \citep[for a thorough discussion, see][]{allen2023economic}.
It follows that the results and interpretation of a SAR model might vary significantly according to the chosen concept of spatial proximity. Moreover, as selecting an appropriate definition that accounts for the key interconnections in the data is often challenging, pinning the model to a single spatial weight matrix may be misleading. To address this issue, recent contributions to the literature have focused on the uncertainty associated with the choice of neighbourhood structures by selecting or combining alternative weight matrices \citep[e.g., see][]{piribauer2016bayesian,debarsy2018flexible}.

Only recently, attention has moved towards spatial models with an unknown weight matrix to be estimated from the data. The main advantage is that the results do not depend on arbitrary distance or neighbour concepts specified a priori; conversely, the inferred matrix would encode the connectivity structure that better explains the relationships in the data. However, this comes at the cost of high dimensional parameter space since the number of free elements in a spatial weight matrix scales quadratically in the length of the cross-section.
Consequently, this new strand of literature builds heavily on statistical techniques for high-dimensional models, such as penalised likelihood approaches \citep{lam2020estimation} or shrinkage priors in a Bayesian setting \citep{krisztin2022bayesian,piribauer2023beyond}. 

Together with the spatial dependence of unknown form, economic and financial data repeatedly observed displays non-negligible temporal dependence over time. An important implication of this stylised fact is that the data might not support the assumption of time-invariant connectivity patterns and strength. Consequently, to account for the dynamics of the unknown dependence structure, we relax this assumption and allow both the weight matrix and the degree of spatial autocorrelation to evolve over time.

This article extends the Bayesian SAR panel model along two important dimensions. First, we relax the typical assumption of an exogenously given spatial weight matrix to make inferences about connectivity patterns. Based on work by \cite{krisztin2022bayesian} and \cite{piribauer2023beyond}, we employ a shrinkage prior to tackling the dimensionality issue and recovering the strongest dependence relationships. Second, we relax the assumption of time-invariant connectivity and degree of spatial autocorrelation. We rely on the features of the multi-sector prices data investigated in the empirical application to design an appropriate mechanism driving the time variation. Distinctive features of many economic and financial data are the cyclical behaviour of the economy at a country-specific level and sporadic abrupt structural changes, which are usually associated with dramatic events, such as financial crises or the propagation of epidemics \citep[e.g., see][]{billio2016interconnections,huber2018markov}. Motivated by these discontinuities in the time series, we propose a Bayesian Markov-switching spatial autoregressive (MS-SAR) panel model to study the impact of covariates and spillover propagation in the different connectivity regimes.

In an empirical application, based on the theoretical framework by \cite{glockerpiribauer2023}, we use the proposed MS-SAR panel model along with monthly data on (three-digit) consumer price index (CPI) subindices for 15 euro-zone countries to estimate the role of consumer preferences in shaping demand-driven cross-price dependencies. The analysis identifies distinct states corresponding to significant economic events and consumer behaviour shifts with marked cross-country variation in spillover dynamics. States characterised by a high network density and widespread inflation spillovers appear particularly during economic transitions, most notably the recent energy price shocks.  

The remainder of this article is organised as follows. Section~\ref{sec:model} proposes a novel SAR model with a time-varying weight matrix, then Section~\ref{sec:estimation} presents the Bayesian approach to inference. The empirical application is discussed in Section~\ref{sec:application}. Finally, Section~\ref{sec:conclusion} concludes.

%Together with the spatial dependence, regional or country-level data repeatedly observed over time displays non-negligible temporal dependence, as is typical in economic and financial data. An important implication of this stylised fact is that the assumption of time-invariant connectivity patterns and strength might not be supported by the data.
%
%Consequently, to account for the dynamics of the unknown dependence structure, we relax this assumption and allow both the weight matrix and spatial dependence parameter to evolve over time.

%We rely on the features of the multi-sector prices data investigated in the empirical application to design an appropriate mechanism driving the time variation. Distinctive features of many economic and financial data are the cyclical behaviour of the economy at a country-specific level and sporadic abrupt structural changes, which are usually associated with dramatic events, such as financial crises or the propagation of epidemics \citep[e.g., see][]{billio2016interconnections,huber2018markov}.
%Motivated by the presence of these discontinuities in the time series, we propose a Markov-switching spatial autoregressive (MS-SAR) model to study the impact of covariates and spillover propagation in the different connectivity regimes.

\section{Model}       \label{sec:model}

We are interested in investigating the direct and indirect propagation of shocks (i.e., spillovers) in a system of $N$ interconnected countries. A standard approach to study these phenomena is a spatial autoregressive (SAR) model \citep[e.g., see][]{anselin1988spatial} with exogenous covariates of the form:
\begin{equation}
\by_t = \rho \bW  \by_t + \bZ_t \bbeta + \boldsymbol{\varepsilon}_t, \qquad \boldsymbol{\varepsilon}_t \sim \mathcal{N}_N(\mathbf{0}, \sigma^2 \mathbf{I}_N),
\label{eq:model_SAR}
\end{equation}
where $\by_t \in \R^N$ denotes an $N$-dimensional vector of country-specific response variables, $\bZ_t \in \R^{N\times M}$ is a full-rank matrix of covariates, with an associated vector of coefficients $\bbeta \in \R^M$. The square matrix $\bW \in \R^{N\times N}$ denotes a spatial weight matrix encoding the contemporaneous connections among the responses, and $\rho \in \R$ is a spatial dependence parameter.

%%%%%%%%%% Unknown W + Temporal dependence %%%%%%%%%%
Most of the literature in spatial statistics considers a time-invariant weight matrix, as in eq.~\eqref{eq:model_SAR}, which is an appropriate assumption for all those settings where $\bW$ represents geographical borders or spatial distance concepts. Conversely, time-invariance is likely violated when the weight matrix is unknown and subject to estimation or summarises a more abstract notion of distance between pairs of subjects, such as in economic and financial applications.
For instance, the Euclidean distance between two collections of economic indicators might be considered a proxy for economic similarity or (inverse) distance between two countries. In cases like this, the temporal evolution of the underlying indicators induces a variation over time in their (economic) distance.
Moreover, when distance metrics other than geographical proximity are considered, little guidance is often provided to the researcher about which criterion best captures the connections among a set of country-level variables. A more agnostic, data-based approach would be a desirable solution to this issue.

Motivated by these key facts, we propose a new model that allows us to infer directly from the data the contemporaneous connectivity structure (or network) among the response variables. Furthermore, to capture the discontinuities in economic and financial time series and account for their time-varying and connectivity structure with persistent links, we introduce a finite state Markov chain driving the dynamics of the pair $(\rho,\bW)$ in eq.~\eqref{eq:model_SAR}.

Following \cite{krisztin2022bayesian}, we assume that each entry $w_{ij,t}$ of the time-varying spatial weight matrix $\bW_t$ is obtained from a hidden spatial binary adjacency matrix $\boldsymbol{\Omega}_t \in \{0,1\}^{N\times N}$ with typical element $\omega_{ij,t}$, that is $\bW_t=f(\bOmega_t)$.
The choice of the linking function $f$ is guided by the restrictions commonly imposed on the weight matrix in static SAR models. First, $\bW_k$ is non-negative with $w_{ij,k}>0$ if variable $j$ is connected (or neighbour) to $i$ in state $k$, and $w_{ij,k}=0$ otherwise. Second, $w_{ii,k}=0$ to avoid self-loops, that is, to prevent a variable from being connected to itself. Third, $\bW_k$ is row-stochastic, meaning that each row sums to unity.
This leads us to the following specification of the link function $f$:
\begin{equation}
w_{ij,t} = \begin{cases}
\omega_{ij,t} / \sum_{j=1}^N \omega_{ij,t} & \text{ if } \sum_{j=1}^N \omega_{ij,t} > 0 \\
0 & \text{otherwise}.
\end{cases}
\label{eq:w_ijt}
\end{equation}
The evolution of $\bOmega_t$ and $\rho_t$ are then modelled using a Markov switching process. Specifically, we assume that $\rho_t$ and each entry $\omega_{ij,t}$ are driven by a common $K$-state hidden Markov chain $\{ \bs_t \}_{t \geq 1}$, that is $\omega_{ij,t} = \omega_{ij,s_t}$ and $\rho_t = \rho_{s_t}$ at each time $t$, where $(\omega_{ij,k}, \rho_k)$, with $k=1,\ldots,K$, are state-specific parameters.
This specification assumes the existence of $K > 1$ states of the world such that at each time $t$, the matrix $\bOmega_t$ (and $\bW_t$) and the dependence parameter $\rho_t$ equal one of $K$ different values.
Finally, by linking the probabilities of connecting any pair $(i,j)$ of variables over time, this construction is able to capture the persistence in the dynamics of the adjacency matrices $\bOmega_t$ and $\bW_t$. This is achieved by means of a time-invariant transition matrix $\bXi = (\bxi_1',\ldots,\bxi_K')'$, where each row $\bxi_k' = (\xi_{k1},\ldots,\xi_{kK})$ is a probability vector and $\xi_{k\ell} = \text{Pr}(s_t=\ell | s_{t-1}=k)$ is the transition probability from state $k$ to state $\ell$.
The resulting Markov switching spatial autoregressive (MS-SAR) model is:
\begin{equation}
\begin{split}
    \by_t & = \rho_{s_t} \bW_{s_t} \by_t + \bZ_t \bbeta + \boldsymbol{\varepsilon}_t, \qquad \boldsymbol{\varepsilon}_t \sim \mathcal{N}_N(\mathbf{0}, \sigma^2 \mathbf{I}_N), \\
    \bW_{s_t} & = f(\bOmega_{s_t}) \\
    P(s_t = k | s_{t-1} = \ell) & = \xi_{\ell,k}, \qquad\qquad \ell,k = 1,\ldots,K.
\end{split}
\label{eq:model_MS-SAR}
\end{equation}

Following \cite{debarsy2022bayesian}, let $\bS_{s_t} = (\mathbf{I}_N -\rho_{s_t} \bW_{s_t})$, then denote with $\bS = \operatorname{blkdiag}(\bS_{s_1},\ldots,\bS_{s_T})$ the $NT\times NT$ block-diagonal matrix whose $t$-th block is the $N\times N$ matrix $\bS_{s_t}$. Moreover, define $\boldsymbol{Y} = (\by_1',\ldots,\by_T')'$, $\boldsymbol{X} = (\bZ_1',\ldots,\bZ_T')'$, $\bOmega = \{ \bOmega_1,\ldots,\bOmega_K \}$, and $\brho = \{ \rho_1,\ldots,\rho_K \}$.
The conditional likelihood can thus be written as:
\begin{equation}
    P(\by | \bs, \bOmega, \brho, \bbeta, \sigma^2) = (2\pi)^{-\frac{NT}{2}} (\sigma^2)^{-\frac{NT}{2}} \abs{\bS} \exp\Big( \! - \frac{1}{2\sigma^2} (\bS \boldsymbol{Y} -\boldsymbol{X}\bbeta)' (\bS \boldsymbol{Y} -\boldsymbol{X}\bbeta) \Big),
\label{eq:likelihood_conditional}
\end{equation}
where $\abs{\bS}$ is the Jacobian of the transformation that depends on the parameters $(\brho,\bOmega)$. Let $N_{k\ell}(\bs)$ count the transitions from state $k$ to state $\ell$, that is $N_{k\ell}(\bs) = \#\{s_{t-1} = k, s_t = \ell \}$, with $\#$ denoting the cardinality of a set. Thus, the complete-data likelihood is:
\begin{equation}
\begin{split}
    P(\by, \bs | \bOmega, \brho, \bbeta, \sigma^2, \bXi) & = (2\pi)^{-\frac{NT}{2}} (\sigma^2)^{-\frac{NT}{2}} \abs{\bS} \exp\Big( \! - \frac{1}{2\sigma^2} (\bS \boldsymbol{Y} -\boldsymbol{X}\bbeta)' (\bS \boldsymbol{Y} -\boldsymbol{X}\bbeta) \Big) \\
    & \quad \times \prod_{k=1}^K \prod_{\ell=1}^K \xi_{k\ell}^{N_{k\ell}(\bs)} P(s_0 | \bXi).
\end{split}
\label{eq:likelihood_complete}
\end{equation}

\section{Bayesian Estimation}   \label{sec:estimation}

\subsection{Prior specification}
In the following, we describe the prior distributions for the state-specific parameters for each state $k=1,\ldots,K$.
Let $\bQ_k = \{ q_{ij,k} \}_{i,j=1}^N$ where $q_{ij,k}$ is the prior probability of a link between $i$ and $j$ in state $k$. We assume the following hierarchical prior distribution for the latent binary spatial matrix $\bOmega_k$:
\begin{align}
    \bOmega_k | \bQ_k & \sim \prod_{i=1}^N \prod_{j=1}^N \mathcal{B}er(\omega_{ij,k} | q_{ij,k}), \\
    q_{ij,k} & \sim \mathcal{B}e(q_{ij,k} | \underline{a}_{q,k}, \underline{b}_{q,k}), \qquad i,j=1,\ldots,N.
\end{align}

The spatial dependence parameter $\rho_k$ in the MS-SAR model in eq.~\eqref{eq:model_MS-SAR} has a key role in determining the overall strength of the spatial relationships.
However, unlike traditional SAR models with known spatial weight matrices, identification issues arise when both the spatial matrix and dependence parameters are inferred from data.
For SAR models with row-standardised known spatial matrix, $\bW$, a sufficient condition to ensure spatial stationarity of the model is $\abs{\rho} < 1$ \citep{lesage2009introduction}. A more relaxed condition is found in \citet[][Lemma 2]{sun1999posterior}, which show that $\rho \in (\lambda_{min}^{-1}, \lambda_{max}^{-1})$ is sufficient, where $\lambda_{min},\lambda_{max}$ denote the minimum and maximum eigenvalues of $\bW$.
Hereinafter, inspired by the common practice, we impose $\rho \in (0,1)$, meaning that only positive spatial autocorrelation is allowed, which is a typical assumption for empirical applications. The most commonly used restriction for known $\bW$ is $\rho \in [0,1)$. However, we rule out $\rho=0$ due to the identification issue that arises in this case when the spatial matrix is inferred from data.
Therefore, we choose a Beta prior for the spatial dependence parameter:
\begin{align}
    \rho_k & \sim \mathcal{B}e(\rho_k | \underline{a}_\rho, \underline{b}_\rho).
\end{align}
Finally, a symmetric Dirichlet prior distribution is assumed for each row of the transition matrix:
\begin{align}
    \bxi_k & \sim \mathcal{D}ir(\bxi_k | \underline{a}_\xi,\ldots,\underline{a}_\xi).
\end{align}

Concerning the remaining non-switching parameters, we assume a Gaussian prior distribution for the coefficient vector and an inverse Gamma prior distribution for the innovation variance:
\begin{align}
    \bbeta & \sim \mathcal{N}_M(\bbeta | \underline{\bmu}_\beta, \underline{\Sigma}_\beta), \qquad\qquad
    \sigma^2  \sim \mathcal{IG}(\sigma^2 | \underline{a}_\sigma, \underline{b}_\sigma).
\end{align}
One may also use an improper prior for $\bbeta$ and $\sigma^2$, that is $P(\bbeta,\sigma^2) \propto 1/\sigma^2$, as the resulting posteriors would be proper \citep[e.g., see][]{debarsy2022bayesian}.

\subsection{Posterior sampling}
To make inferences on the model parameters, we design an efficient Gibbs sampler targeting the joint posterior distribution that cycles over the following steps:\footnote{We refer to the Supplement for the derivation of the full conditional posterior distributions.}
\begin{enumerate}
    \item Draw the path of the latent states from $[\bs | \by, \bXi, \bOmega, \brho, \bbeta, \sigma^2]$ using a forward-filtering backward-sampling (FFBS) algorithm.
    \item Draw the spatial dependence parameters from $[\rho_k | \by, \bs, \bOmega, \bbeta, \sigma^2]$ using a Griddy-Gibbs algorithm.
    \item Draw the entries of the latent binary matrix from $[\omega_{ij,k} | q_{ij,k}, \by, \bs, \rho_k, \bbeta, \sigma^2]$.
    \item Draw the probability of each link in $\bOmega_k$ from
    \begin{equation}
    q_{ij,k} | \omega_{ij,k} \sim \mathcal{B}e\big( q_{ij,k} | \underline{a}_{q,k} + \omega_{ij,k}, \, \underline{b}_{q,k} + (1-\omega_{ij,k}) \big).
    \end{equation}
    \item Draw the rows of the transition matrix from
    \begin{equation}
    \bxi_k | \bs \sim \mathcal{D}ir\big( \bxi_k | \underline{a}_{\xi,k} + N_{k1}(\bs),\ldots,\underline{a}_{\xi,k} + N_{kK}(\bs) \big).
    \end{equation}
    \item Draw the covariates' coefficients from 
	\begin{equation}
	\bbeta | \by, \bs, \bOmega, \brho, \sigma^2 \sim \mathcal{N}_M\big( \bbeta | \overline{\bmu}_\beta, \, \overline{\Sigma}_\beta \big).
	\end{equation}
    \item Draw the structural innovations variance from
	\begin{equation}
	\sigma^2 | \by, \bs, \bOmega, \brho, \bbeta \sim \mathcal{IG}\Big( \sigma^2 \big| \underline{a}_\sigma + \frac{T}{2}, \, \underline{b}_\sigma + \frac{1}{2} \sum_{t=1}^T (\bS_{s_t}\by_t -\bZ_t \bbeta)'(\bS_{s_t}\by_t -\bZ_t \bbeta) \Big).
	\end{equation}
\end{enumerate}
%
%\begin{enumerate}
%    \item draw $[\bs | \by, \bXi, \bOmega, \brho, \bbeta, \sigma^2]$ using a forward-filtering backward-sampling (FFBS) algorithm;
%    \item draw $[\rho_k | \by, \bs, \bOmega, \bbeta, \sigma^2]$ using a Griddy-Gibbs algorithm;
%    \item draw $[\omega_{ij,k} | q_{ij,k}, \by, \bs, \rho_k, \bbeta, \sigma^2]$ from a Bernoulli distribution;
%    \item draw $[q_{ij,k} | \omega_{ij,k}]$ from a Beta distribution;
%    \item draw $[\bxi_k | \bs]$ from a Dirichlet distribution;
%    \item draw $[\bbeta | \by, \bs, \bOmega, \brho, \sigma^2]$ from a Gaussian distribution;
%    \item draw $[\sigma^2 | \by, \bs, \bOmega, \brho, \bbeta]$ from an inverse Gamma distribution.
%\end{enumerate}
%
Steps 4-5-6-7 are standard, and we refer to the Supplement for full computational details.
The entire path of the latent states is sampled in Step 1 by using a forward-filtering backward-sampling (FFBS) algorithm \citep{fruhwirth2006finite}. This algorithm consists of a forward pass to compute the filtered probabilities of each state and a backward pass to compute the smoothed probabilities and simulate the entire trajectory of $\bs$ jointly, which is more efficient than drawing each $s_t$ from its full conditional distribution.

The dependence parameter has a complex full conditional distribution due to the complex way it enters the likelihood, for which no conjugate prior is available. However, since $\rho_k$ is a scalar parameter distributed on a bounded support, several methods for approximating its posterior distribution are available. In particular, as in \cite{lesage2009introduction}, we rely on the Griddy-Gibbs algorithm \citep{ritter1992facilitating} that consists in discretising the support $(0,1)$ of $\rho_k$, and sampling from the re-normalised approximated discrete distribution.

Finally, Step 3 samples the entries of the latent binary matrix $\bOmega_k$ relying on the efficient algorithm proposed in \cite{krisztin2022bayesian}. Denoting with $\mathbf{Y}^k = \{ \by_t : s_t = k\}$ the observations allocated to the $k$th state and with $\bOmega_{-ij,k}$ all the elements of $\bOmega_k$ except $\omega_{ij,k}$, the not normalised posterior full conditional probabilities are given by
\begin{align*}
\tilde{p}_{ij,k}^{(1)} & = P(\omega_{ij,k} = 1 | \bOmega_{-ij,k}, \by, \bs, \bbeta, \rho_k, \bsigma^2, q_{ij,k}) \\
    & \propto q_{ij,k} \prod_{t\in\mathcal{T}_k} \abs{S_{k,1}} \exp\Big( -\frac{1}{2} (\bS_k \by_t -\bZ_t\bbeta)' \Sigma^{-1} (\bS_k \by_t -\bZ_t\bbeta) \Big), \\
\tilde{p}_{ij,k}^{(0)} & = P(\omega_{ij,k} = 0 | \bOmega_{-ij,k}, \by, \bs, \bbeta, \rho_k, \bsigma^2, q_{ij,k}) \\
    & \propto (1-q_{ij,k}) \prod_{t\in\mathcal{T}_k} \abs{\bS_{k,0}} \exp\Big( -\frac{1}{2} (\bS_k \by_t -\bZ_t\bbeta)' \boldsymbol{\Sigma}^{-1} (\bS_k \by_t -\bZ_t\bbeta) \Big),
\end{align*}
where $S_{k,1}$ and $S_{k,0}$ are obtained from $S_k$ by modifying the spatial weight matrix $\bW_k$ via setting $\omega_{ij,k} = 1$ and $\omega_{ij,k} = 0$, respectively. Therefore, each element $\omega_{ij,k}$ has a Bernoulli posterior full conditional distribution $\mathcal{B}ern(\omega_{ij,k} | \overline{q}_{ij,k})$, where $\overline{q}_{ij,k} = \tilde{p}_{ij,k}^{(1)} / (\tilde{p}_{ij,k}^{(0)} + \tilde{p}_{ij,k}^{(1)})$.

The selection of the number of states in the model, $K$, is performed using the DIC$_5$ proposed in \cite{celeux2006deviance}, which is a modified version of the deviance information criterion \citep[DIC, see][]{spiegelhalter2002bayesian}.

\section{Empirical application}   \label{sec:application}

Based on theoretical underpinnings in \cite{glockerpiribauer2023}, we apply the proposed MS-SAR model to illustrate the role of consumer preferences in shaping demand-driven cross-price dependencies by using monthly data on three-digit CPI subindices for 15 euro-zone countries.

\subsection{Preferences, demand and cross price dependencies\label{sec:econ_teory}}

Cross-demand describes the relationship between the change in demand for a good or service in response to a change in the price of a related good or service. An important aspect of this theory is that a change in the price of good $i$ will, in turn, shift the demand curve, causing both the demand for good $j$ and its price to change. Therefore, given either a substitutability or complementarity relationship, a change in the price of good $i$ will not only change the demand for good $j$ but also the price of good $j$. Thus, changes in the price of one good can affect the prices of other goods through the demand system. 

\cite{glockerpiribauer2023} establish a relationship between demand-driven cross-price dependencies and the consumer price index (CPI, $p$), and hence the CPI inflation rate ($\pi$). In what follows, we briefly discuss this. Consider $N$ goods and let $x_i(p_i,\bp_{-i},m)$ be the Marshallian demand function of good $i$ as a function of its own price $p_i$, the vector of the prices of all other goods $\bp_{-i}$, and income $m$. We collect the corresponding inverse demand functions in the $N$-dimensional vector-valued function $\bp^f$. Assuming perfectly competitive markets and price inelastic supply, the vector $\bp^f$ represents equilibrium prices. Computing the total differential, the change $\Delta \bp^f$ implies that 
\begin{equation}
\Delta \bp^f = \bA \Delta \bp^f + \mathbf{u},
\label{eq:delta_pf}
\end{equation} 
where the vector $\mathbf{u}=\nabla_{m} \bp^f(\bp,\bx,m) \cdot \Delta m$ which captures income-demand effects. Most importantly, $\bA$ is a Jacobian matrix of first derivatives, that is $\bA = \nabla_{\bp} \bp^f$, which contains cross-price interdependencies $\partial p_i/\partial p_j$, $\forall\ i\neq j$ with $i,j=1,\ldots,N$ (and zeros on the main diagonal). The cross-price dependencies measure the extent to which the price of good $i$ changes as a result of a change in the price of good $j$ within the demand system. Since cross-price effects arise due to the substitutability or complementarity of goods (and services), cross-demand, and hence the non-zero entries in the Jacobian matrix $\bA$, can be either positive or negative. 

% \todo{M: $\bw$ seems too similar to the matrix $\bW$...maybe better to use $\bomega$ or $\bq$}

To obtain a relationship between the Jacobian matrix $\bA$ and the CPI inflation rate\footnote{We approximate the geometric mean, as used to construct the CPI, by the arithmetic mean.}, let the CPI $p$ be given by $p = \bomega'\bp^f$, where the $N$-dimensional column vector $\bomega$ contains the weights (expenditure shares) of each good. Using this expression, if the weights $\bomega$ are fixed at some constant level, then the change in the CPI is given by $\Delta p = \bomega' \Delta \bp^f$ and the CPI inflation rate is given by the relative change in the CPI $\pi = \Delta p/p$. Substituting eq.~\eqref{eq:delta_pf} into the latter expression yields the following relationship between prices shape the CPI inflation rate:
\begin{equation}
    \pi = \bomega' (\mathbf{I}_n-\bA)^{-1} \tilde{\mathbf{u}},
\label{eq:pi_links}
\end{equation}
where $\tilde{\mathbf{u}} = \mathbf{u}/p$. For example, if the Jacobian $\bA$ is the null matrix, then the effect of a shock to a single price on the CPI inflation rate is equal to that price's weight times the shock's size. Instead, in the presence of cross-price dependencies, the Jacobian $\bA$ is nonzero, implying that a shock to price $p_i$ can affect the CPI inflation rate both directly, via the price $p_i$ and indirectly, via the spillover effects arising from the cross-price dependencies.

To keep the exercise tractable, we restrict the analysis to non-negative entries in the Jacobian matrix, thus implying that all links between prices arise from complementarities between goods (and services). While this is a strong assumption and is likely to overestimate the impact of an exogenous shock to a given price on CPI inflation, it is justified by the fact that \cite{Regmi2010} and \cite{glockerpiribauer2023}, for example, emphasise the dominance of complementarity over substitution relationships.

\subsection{The data\label{eq:data}}

The goods and services in the CPI basket are classified by purpose into consumption groups (COICOP) at different levels of (dis)aggregation called \emph{level of structure} (two-digit, three-digit, etc.). We use the three-digit COICOP structure of the CPI, which consists of 44 different monthly price sub-indices.
We exclude sub-indices for which the sample does not extend back to at least 2002. These are the price sub-indices related to educational goods and services (cp\_101 to cp\_105), leaving a total of $N=39$ sub-indices and their respective basket weights.\footnote{Weights have been adjusted to take account of the omission of five sub-indices.}
Details of the sub-indices are given in the Supplement. We collect data for 15 euro-zone countries (Austria, Belgium, Cyprus, Estonia, Finland, France, Germany, Greece, Ireland, Italy, Latvia, Lithuania, the Netherlands, Portugal and Spain). 

For each country in our sample, we consider the monthly relative year-over-year changes from January 2002 to June 2023, resulting in $T=246$ time observations. Besides these CPI-based data and additional monthly variables as control variables, explained in detail below, we also use the weights ($\bomega$) for each price sub-index of the three-digit COICOP structure for the subsequent analysis.

\subsection{From theory to an econometric specification\label{sec:econ2metrix}}

We apply the proposed MS-SAR model to investigate the cross-dependencies between the aforementioned CPI-based price sub-indices at the 3-digit level. In this context, we aim to estimate a version of equation \eqref{eq:delta_pf}.
We apply the MS-SAR panel model in eq.~\eqref{eq:model_MS-SAR} where the dependent variable $\by_t = \Delta \bp^f_t$ is an $N$-dimensional vector comprising monthly information on the $N=39$  price sub-indices and expressed as year-over-year (relative) change. The MS-SAR model is applied individually to all countries in our sample. Therefore, the \textit{spatial} element refers only to the individual price sub-indices and not to geographical regions. We include item-fixed effects in each country model to control for item-specific heterogeneity.

Significantly, changes in consumer behaviour can affect cross-price dependencies along two dimensions. The first concerns the amount of nonzero elements in $\bA$ and, hence, the amount of other goods' prices being affected once the price of a particular good changes. We control for this effect by considering a state-dependency in the matrix, $\bW_{s_t}$. The second concerns the extent of inertia. Once a price shock propagates through the demand system, its effect could either  abate quickly or not. We consider a  state-dependency in the scalar, $\rho_{s_t}$, to account for this matter. In both cases, the state-dependency represents changes in consumer preferences and, hence, consumer behaviour.

%The use of inverse demand functions to estimate demand-driven cross-price dependencies leads to an identification issue arising since prices are the result of the interplay between supply and demand, leading to equilibrium outcomes under market clearing conditions. In cases where the supply is price inelastic (such as administered prices or short-term supply), endogeneity is less of an issue. However, this is generally not the case and a method to hold demand constant while adjusting supply is required. Therefore, we include a wide range of supply-side variables in the explanatory variable matrix ($\mathbf{Z}_t$) that primarily affect price and indirectly affect demand. We can isolate the demand-driven cross-price interdependencies within the $\bW_{s_t}$ matrix by controlling for these supply-side factors.

The explanatory variables, $\mathbf{Z}_t$, include several producer price sub-indices\footnote{This serves to identify demand and hence to solve the endogeneity problem related to estimating demand and supply functions. We can isolate the demand-driven cross-price interdependencies within the $\bW_{s_t}$ matrix by controlling for these supply-side factors.} (e.g., food, textiles, energy), administered price measures, the nominal effective exchange rate, and global indicators such as shipping costs \citep{CARRIERESWALLOW2023102771}, Brent crude oil prices, and natural gas prices (Dutch TTF). In addition, we include the overall CPI inflation rate as a control variable. This choice is based on the common industry practice of adjusting prices based on past CPI inflation rates (inflation indexation of prices). These variables are used for each country in our analysis.

\subsection{Results}

%We estimate the MS-SAR model individually for each country. 
Given the abundance of empirical output, we organise our discussion as follows: Section~\ref{sec:ctry_res} provides country-specific evidence, focusing on two specific countries to illustrate how the model works in the context of their unique characteristics. Section~\ref{sec:allctry_res} provides consolidated statistics and discussion, where we attempt to synthesise the results from all countries in our sample.

To analyse the impact of demand shocks on CPI inflation, it is helpful to rewrite the model in its reduced form
\begin{equation}
    \by_t = (\mathbf{I}_N-\rho_{s_t} \bW_{s_t})^{-1} (\bZ_t \bbeta + \boldsymbol{\varepsilon}_t),
    %\Delta \boldsymbol{p}^f_t = (\boldsymbol{I}_N-\rho_{s_t} \bW_{s_t})^{-1}(\boldsymbol{Z}_t\boldsymbol{\beta}+\boldsymbol{\varepsilon}_t),
\end{equation}
where the $N\times N$ matrix $\bS_{s_t} \coloneqq (\mathbf{I}_N - \rho_{s_t} \bW_{s_t} )^{-1} = \mathbf{I}_N + \rho_{s_t}\bW_{s_t} + \rho_{s_t}^2\bW_{s_t}^2 + \cdots$ is commonly referred to as the spatial multiplier matrix. It describes the transmission of demand shocks $\boldsymbol{\varepsilon}_t$ (and thus of the elements in $\mathbf{Z}_t$) in the system of cross-price dependencies. The elements on the diagonal of the spatial multiplier matrix $\bS_{s_t}$, called direct effects in the spatial econometrics literature \citep[e.g., see][]{lesage2009introduction}, contain information about the impact of a shock $\varepsilon_{i,t}$ on the price sub-index $p_i$, namely the own-price effect.
The Neumann series expansion of the spatial multiplier matrix shows that these main diagonal elements already contain feedback effects due to cross-price dependencies. The off-diagonal elements in $\bS_{s_t}$ comprise the indirect (or spillover) effects and capture the impact of a shock $\varepsilon_{i,t}$ on other price sub-indices $p_j$, $\forall\ j\neq i$. 

Motivated by the interest in analysing the impact of shocks on the overall CPI inflation rate $\pi$, we exploit eq.~\eqref{eq:pi_links} to relate the direct and indirect effects of shocks on particular price sub-indices to the overall CPI inflation rate.
Let us first define the $N$-dimensional vector $\boldsymbol{\delta}_t$, which collects the direct effects on the CPI inflation rate at time $t$, as 
\begin{equation}
    \boldsymbol{\delta}_t = \bomega'_t \operatorname{diag}(\bS_{s_t}),
\label{eq:diri}
\end{equation}
where the $i$-th element $\delta_{i,t}$ captures the direct effect of a shock to the $i$-th price subindex on the CPI inflation rate and $\operatorname{diag}(\bS_{s_t})$ denotes the diagonal matrix operator such that $\operatorname{diag}(\bS_{s_t})$ represents an $n \times n$ diagonal matrix. Instead, the contribution of the spillover (or indirect) effects to the CPI inflation rate is given by the off-diagonal elements of the spatial multiplier
\begin{equation}
    \boldsymbol{\zeta}_t = \bomega_t' \big( \bS_{s_t}-\text{diag}(\bS_{s_t}) \big).
\label{eq:spilli}
\end{equation}
Indirect effects measure the impact of shocks to a particular price sub-index on the CPI inflation rate resulting only from the interaction with other price sub-indexes.
%That is, if the matrix $\bW_{s_t}$ is equal to the zero matrix, then all spillover effects are equal to zero.
Finally, the total effect combines both direct and spillover effects and is given by $\boldsymbol{\tau}_t = \bomega_t' \bS_{s_t}$.
%
%\begin{equation}
%    \boldsymbol{\tau}_t = \bomega_t' \bS_{s_t}.
%\label{eq:totaleff}
%\end{equation}
%
In the following, we will focus on the extent to which direct and indirect effects vary across countries and, most importantly, across the states of the world as captured by the Markov chain.

\subsubsection{Country-specific results\label{sec:ctry_res}}

\begin{figure}[ht!]
\caption{Smoothed state probabilities}
\label{fig:TPfig}
\setcounter{subfigure}{0}
\centering
\subfigure[Greece]{
\includegraphics[width=\textwidth]{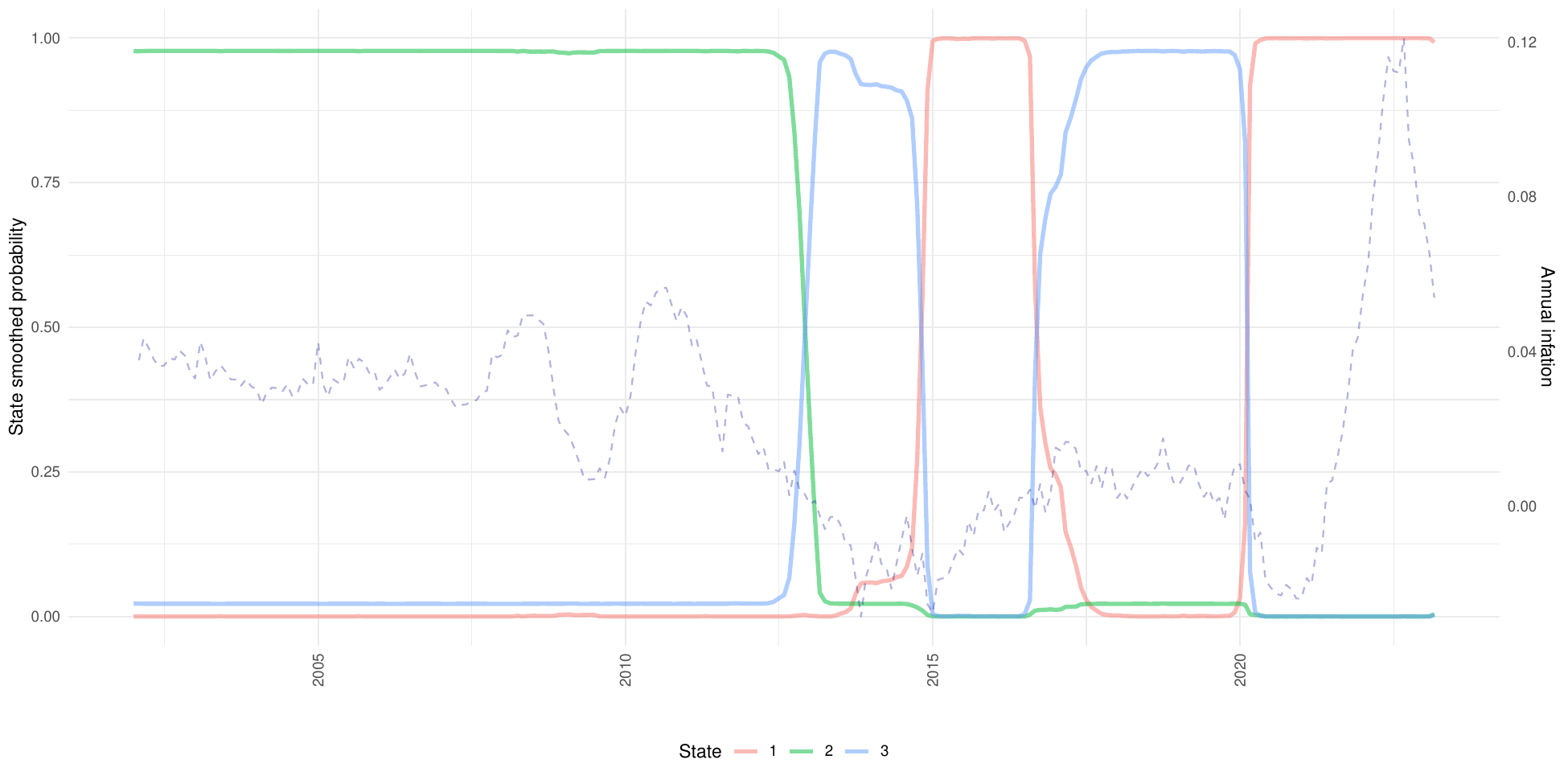}
\label{fig:ELresu1}
}
\subfigure[Germany]{
\includegraphics[width=\textwidth]{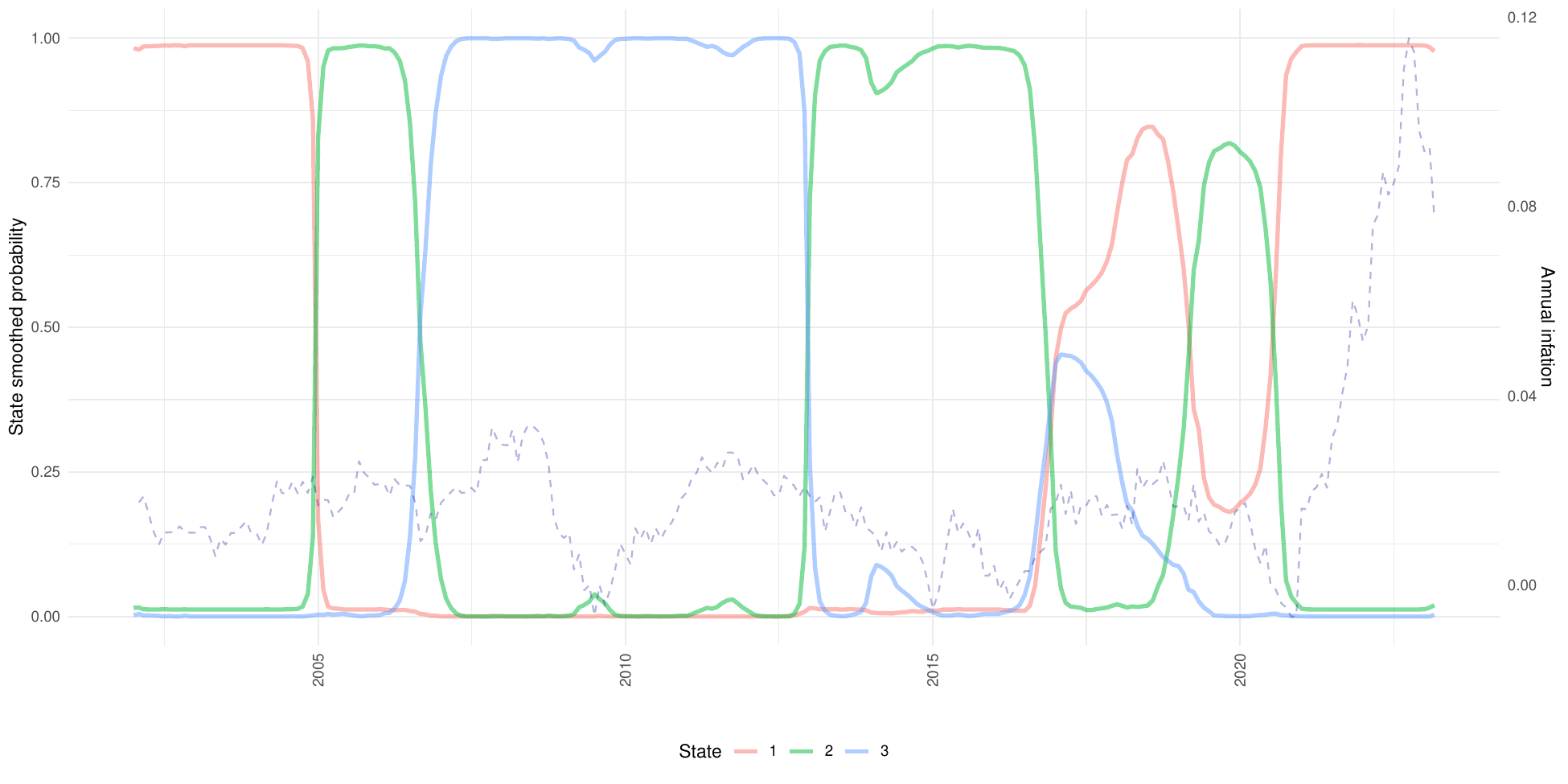}
\label{fig:DEresu1}
}
\fnote{\scriptsize Note: The Figure shows the time trajectory of the (smoothed) state probabilities for Greece (upper panel) and Germany (lower panel). The probabilities for states 1, 2 and 3 are shown in red, green, and blue colors, respectively. The dashed line depicts the path of the respective  CPI inflation rate.}
\end{figure}

We present a detailed analysis of Greece (EL) and Germany (DE), chosen for their interesting patterns, which provide valuable insights into the functioning and understanding of our empirical model.\footnote{The results of other countries in the sample are too numerous to be included here, but are available upon request to the authors.}
We use the DIC criterion to determine the number of hidden states of the Markov chain, resulting in $K=3$ for both countries. The two panels in Figure ~\ref{fig:TPfig}  report the posterior estimate of the (smoothed) state probabilities for Greece (upper panel) and Germany (lower panel). In each case, we compare the time trajectory of the (smoothed) state probabilities to the overall CPI inflation rate (dashed line in the panels).

The results show several state transitions for Greece. From a macroeconomic perspective, the first coincides with the Greek debt crisis in 2012. The subsequent state transitions resemble significant energy price changes, the first of which occurred in 2014/2015 with a considerable drop in oil prices, followed by a subsequent increase in 2017. The final state transition coincides with the outbreak of the COVID-19 pandemic in early 2020.
Regarding Germany, it is noteworthy that the state transitions reflect either economic crises or their recovery. For example, the first (2004/2005) corresponds to a country-specific episode of economic slack, while the second (2007) corresponds to the subsequent recovery. The third (2013) corresponds to the recovery from the European debt crisis, while the last partly reflects the energy price hike and, thus, the recovery from the COVID-19 recession. Between 2017 and 2020, the smoothed probabilities are very close to each other, thus suggesting no clear evidence in favour of any particular state.

Given our focus on estimating (inverse) consumer demand functions, the evolution in state probabilities indicates changes in consumer preferences and, hence in the composition of household expenditure, which in turn leads to changes in cross-price dependencies. This adjustment is particularly pronounced during switches of economic conditions, from boom to bust and vice versa, indicating sudden and dramatic changes in demand and, hence in cross-price dependencies.
This phenomenon was particularly evident during the COVID-19 pandemic and may have been exacerbated by policy-related closure measures: the enforced closure of physical retail outlets led to a significant shift towards online shopping. \cite{gorodnichenko2017price} and \cite{gorodnichenko2018price} underline that the shift to online retailing had a significant impact on price dynamics, ultimately leading to changes in demand-driven cross-price dependencies and hence the overall CPI inflation rate.\footnote{As the matrix $\bW_{s_t}$ represents a directed weighted network, the presence of cycles between nodes can be important. A cycle involving only two nodes (i.e., a directed link from node $i$ to $j$ and another from $j$ to $i$) can have a large overall effect even though there are no other edges in the network. We refer the interested reader to \cite{fan2021} and \cite{glockerpiribauer2023} for further discussion.}

\begin{table}[h!t]
\centering
\caption{Estimated network statistics}
\label{tab:netwtopo}
\makebox[1 \textwidth][c]{
\begin{threeparttable}
\begin{footnotesize}
\begin{tabular}{c cccccc}
\hline \\[-5pt]
	& \multicolumn{2}{c}{State No.1} & \multicolumn{2}{c}{State No.2} & \multicolumn{2}{c}{State No.3} \\
	& \multicolumn{2}{c}{\rule[2mm]{2.5cm}{0.2mm}} & \multicolumn{2}{c}{\rule[2mm]{2.5cm}{0.2mm}} & \multicolumn{2}{c}{\rule[2mm]{2.5cm}{0.2mm}} \\
	&	DE	& 	EL	&	DE	& 	EL	&	DE	& 	EL	\\
 \hline \\[-5pt]
    & \multicolumn{6}{c}{Network statistics} \\
    & \multicolumn{6}{l}{\rule[2mm]{8cm}{0.2mm}} \\
    Link density (in \%)	&	3.10	&	3.24	&	1.48	&	2.70	&	1.21	&	2.02	\\
    Network density ($\bW_{s_t}$) & 1.00 & 1.17 & 0.59 & 0.98 & 0.51 & 0.84 \\ 
    Network density ($\rho_{s_t} \bW_{s_t}$) & 0.47	& 0.70 & 0.19 &	0.33 & 0.09 & 0.25 \\
\hline \\[-5pt]
    & \multicolumn{6}{c}{Spatial auto-regressive parameter} \\ 
    & \multicolumn{6}{l}{\rule[2mm]{8cm}{0.2mm}} \\
   $\rho_{s_t}$ post. mean & 0.47 & 0.60 & 0.33& 0.34 & 0.18 & 0.29 \\
   $\rho_{s_t}$ post. std. dev. & 0.03 & 0.02 & 0.03& 0.01 & 0.02 & 0.02 \\
\hline
\end{tabular}
\end{footnotesize}
\begin{scriptsize}
\begin{tablenotes}[para]
The values in the table are multiplied by 100 in each case. The link density is the ratio between the number of links (edges) and the maximum possible number of links in the network. The network density is the average value of the size of the existing links. It extends the link density by considering the numerical value associated with each link (in a weighted network). Hence, the network density measures the strength effect for a given link density. 
\end{tablenotes}
\end{scriptsize}
\end{threeparttable}
}
\end{table}

The changes of states, the switch in the probability of a particular state relative to the others, also lead to significant changes in the structure of demand-driven cross-price dependencies. Details are provided in Table~\ref{tab:netwtopo} using two commonly used measures of network topology: link density \citep{Newman2010} and network density \citep{horvath2011weighted}. The former is the ratio between the number of links and the maximum possible number of links in the network. The network density, defined as the average value of the size of the existing links, extends the link density by accounting for the weight associated with each link in a weighted network; as such, it measures the average strength of the links. Both connectedness measures quantify the spillover effects of shocks to individual price sub-indices. In computing these network statistics, only those entries in the matrix $\bW_{s_t}$ whose posterior inclusion probabilities exceed a value of $0.68$ are considered; those with a value below this threshold are set to zero.\footnote{This choice is motivated by the fact that the proposed model can shrink the entries of $\bW_{s_t}$ toward zero but does not allow for exact zeros. Therefore, we follow common practice in Bayesian shrinkage estimation and set a hard threshold to obtain zeros in the estimated adjacency matrices.}

In both countries under scrutiny, state $1$ is characterised by the most significant number of links with a concomitantly high overall network density. In contrast, state $3$ is associated with the lowest degree of interconnectedness. Looking at the estimated path of the state probabilities, we find the transition to state $1$ in conjunction with dramatic changes in the overall CPI inflation rate resulting from changes in (aggregate) demand. Since state $1$ reflects a situation of high linkage and network density, small shocks to the demand for particular goods (i.e., individual price sub-indices) quickly spill over to many other prices, eventually leading to a broad-based price (and hence inflationary) impact. Given that we consider a framework where all the links are positive-valued, any additional link exacerbates the transmission mechanism of shocks, thus increasing the likelihood of the previously mentioned phenomenon being particularly pronounced.

It is noteworthy that the estimated spatial weight parameter significantly differ across the states, as shown in the bottom panel of Table~\ref{tab:netwtopo}. The posterior mean is highest in state 1 and lowest in state 3, reinforcing the previous claim that these states represent periods of strong and weak spatial dependence, respectively. Instead, the posterior standard deviation is small and similar across all the states.
These findings support the claim that both the structure of the spatial relationships and their strength vary over time, which is the key motivation underlying the Markov-switching specification for the pair $(\rho,\bW)$ in model \eqref{eq:model_MS-SAR}.
Interestingly, our results confirm the claim recently made by \cite{scida2023structural}, which performed a rolling window analysis of a SAR model with time invariant parameters. In contrast, the proposed MS-SAR model is explicitly designed to capture the switching values of the parameter and network structure.

\begin{figure}[h!t]
\caption{Direct, spillover and total effects}
\label{fig:SPfig}
\setcounter{subfigure}{0}
\centering
\subfigure[Greece]{
\includegraphics[height=0.37\textheight,width=\textwidth]{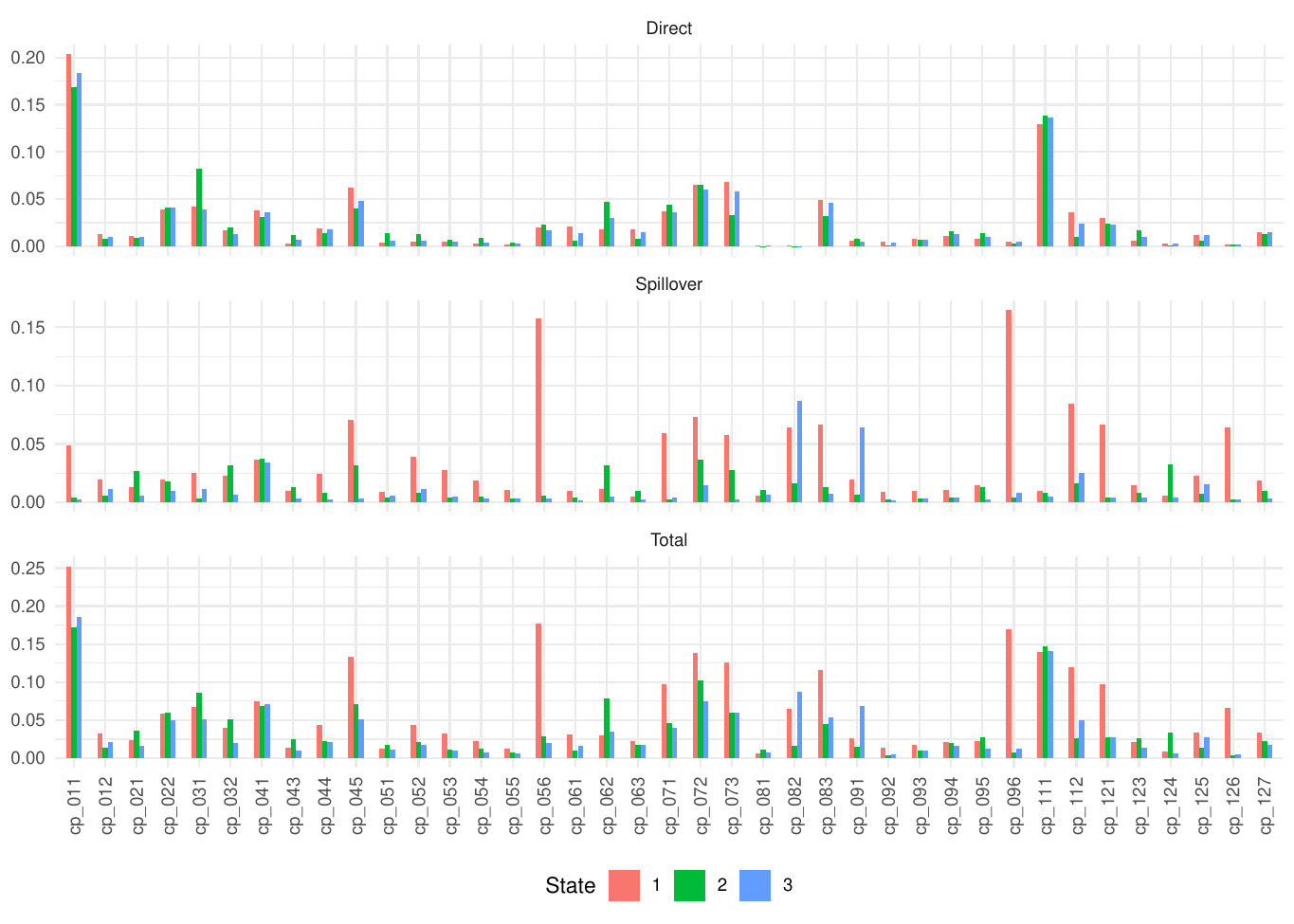}
\label{fig:ELresu2}
}
\subfigure[Germany]{
\includegraphics[height=0.37\textheight,width=\textwidth]{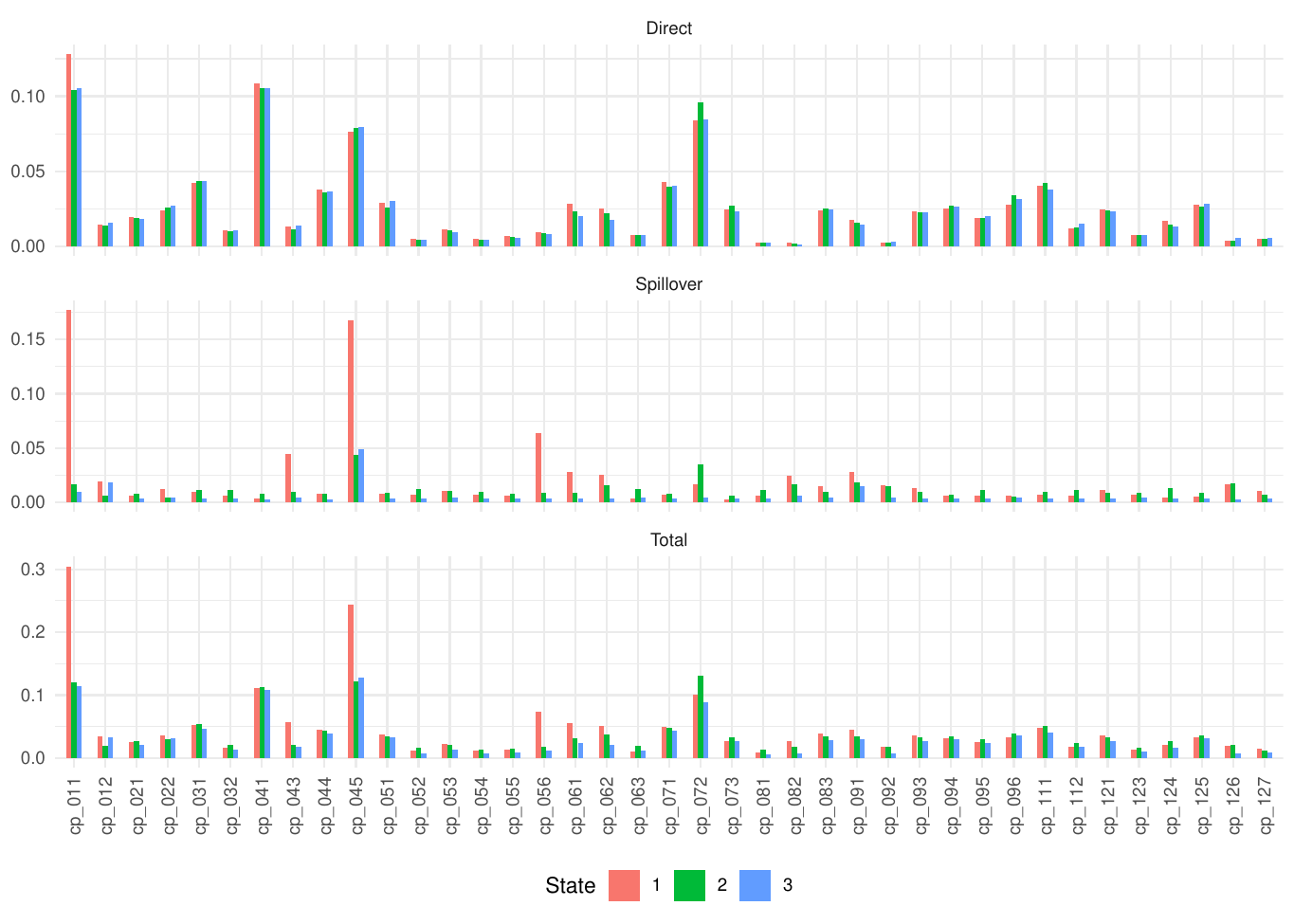}
\label{fig:DEresu2}
}
\fnote{\scriptsize Note: The figure shows the posterior means of the direct ($\boldsymbol{\delta}_t$), spillover ($\boldsymbol{\zeta}_t$), and total effects ($\boldsymbol{\tau}_t$) for the different states and CPI subindices. The upper panel depicts the results for Greece and the lower one for Germany.}
\end{figure}

The dominance of state $1$ is also apparent when considering the effects of shocks to individual price sub-indices on the CPI inflation rate. To this end, we now consider the decomposition of the impact of a shock to each single price sub-index on the CPI inflation rate into direct and indirect effects, as motivated by eq.~\eqref{eq:diri} and \eqref{eq:spilli}.
The two panels in Figure~\ref{fig:SPfig} show the results for each price sub-index across all identified states for both countries. The calculations are based on the average over time of the weights $\bomega$. Specifically, we first compute the posterior mean of each state-specific spatial multiplier matrix $\bS_k$. For each period $t$ where the posterior inclusion probability of state $k$ was greater than $0.5$, we then calculate the direct, indirect, and total effects using the corresponding weights $\bomega_t$. Finally, we average these effects over time.

The results for both countries suggest that, in many cases, the most considerable spillovers (middle sub-panel) are associated with state 1 (red bar). In contrast, the other two states are associated with significantly smaller spillovers in magnitude.
Another interesting result in this context concerns the importance of certain price sub-indices in shaping the overall CPI inflation rate.

For instance, the results for Greece show that, in some cases, a large overall effect is due solely to strong direct effects (e.g., for the categories ``food'' and ``restaurant services''); that is, the corresponding price indices have a large (average) weight in the CPI, while in other cases a large overall effect is due to significant spillover effects (e.g., for the categories ``Package holidays'' and ``Telephone and Telefax Equipment''). 
A similar pattern is observable for Germany, though the price sub-indices shaping the spillover effects most considerable in size are instead ``Electricity, Gas and Other Fuels'' and ``Goods and Services for Routine Household Maintenance'' apart from the category ``Food'' whose large total effect is established by both large direct and spillover effects.

\FloatBarrier
\subsubsection{Evidence across countries\label{sec:allctry_res}}

\begin{sidewaysfigure}[h!]
    \centering
    \includegraphics[scale=0.80]{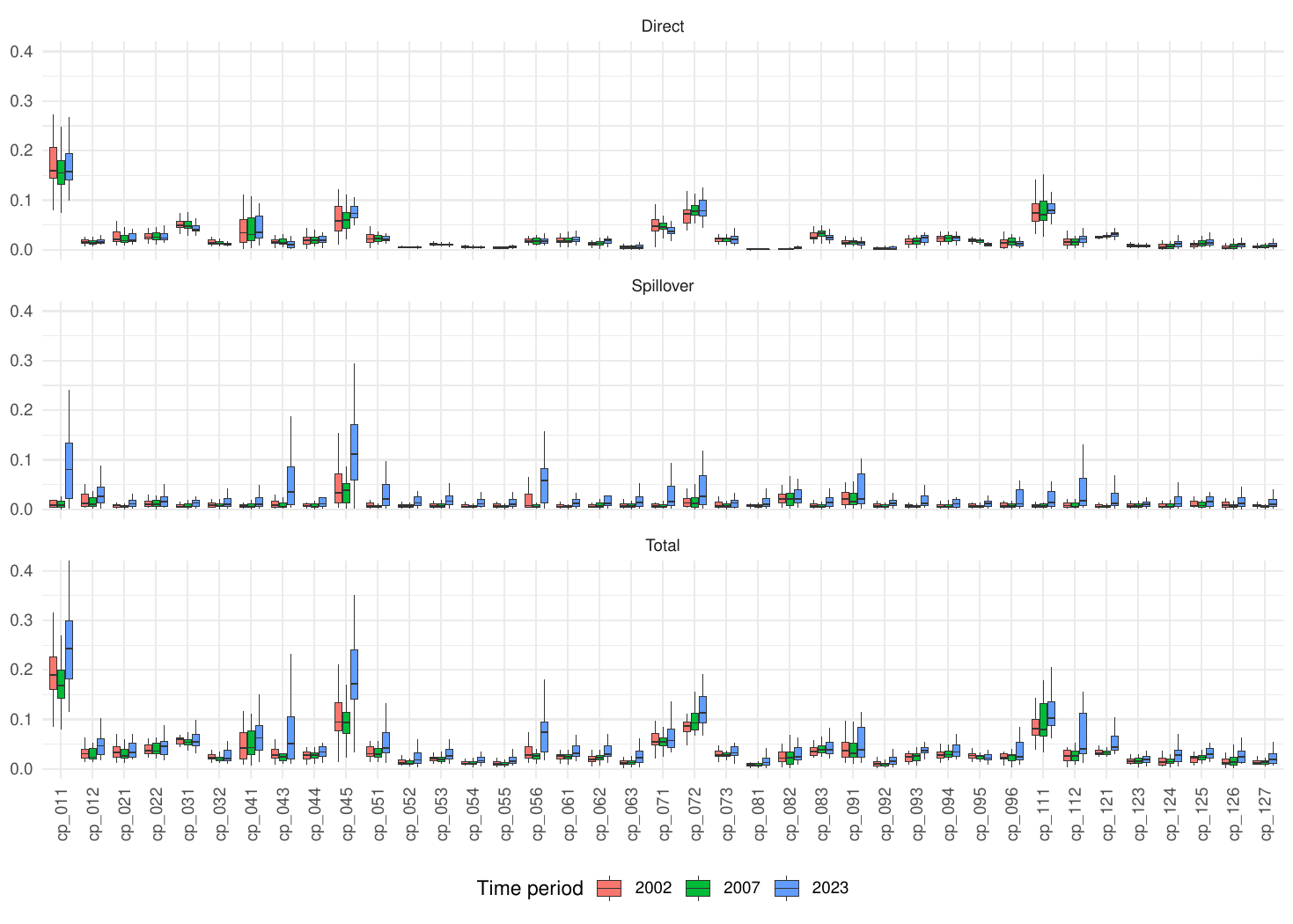}
    \caption{Comparing direct, spillover and total impacts among countries for different periods}
    \label{fig:boxplot_impacts}
    \fnote{\scriptsize Note: The figure shows posterior mean estimates for direct ($\boldsymbol{\delta}_t$), spillover ($\boldsymbol{\zeta}_t$), and total effects ($\boldsymbol{\tau}_t$) for the CPI subindices under scrutiny. For each subindex, the figure depicts grouped boxplots showing the distribution of the impacts of the countries in the Eurozone for the periods 2002 (red), 2007 (green), and 2023 (blue).}
\end{sidewaysfigure}

As the previous country-specific results highlight some form of country heterogeneity, we now examine the cross-country variation in more detail, using all 15 countries in the sample.
Figure~\ref{fig:boxplot_impacts} provides a comprehensive overview of the effects of different price sub-indices on the CPI inflation rate for three selected years, which represent different phases that are essentially common to all countries in our sample: the beginning of an expansionary period (2002), the end of a boom (2007), and the peak of the energy crisis in 2022/2023 (2023).
The figure breaks down the impacts along three dimensions: (i) direct, (ii) indirect, and (iii) total impacts and the bars measure the extent of country variation along these three dimensions for each of the selected years.

%\begin{landscape}
%\begin{figure}
%\end{figure}
%\end{landscape}

One striking result is the significant cross-country variation in some price sub-categories, such as ``Electricity, Gas \& Other Fuels'', ``Operation of Personal Transport Equipment'', and ``Other Recreational Items \& Equipment, Gardens \& Pets''. In each case, the total effects show considerable cross-country variation, mainly due to cross-country differences in spillover effects. On the other hand, the direct effects are relatively similar in size across countries, implying small bars on average. 

Interestingly, the estimated direct effects are relatively constant between 2002 and 2007, whereas the spillover effects show more time variation. However, in 2023, the time variation becomes more pronounced compared to the other years. This dramatic change goes beyond energy-related prices and is particularly pronounced in categories such as ``Accommodation Services'' and ``Goods \& Services for Routine Household Maintenance''. Crucially, the variation in total effects in 2023 compared with 2002 and 2007 is mainly due to changes in spillovers rather than weight changes. The slight change of direct impacts over time is primarily because the weights of the price sub-indices in the CPI are only marginally adjusted. The comparatively high importance of spillovers thus underlines the importance of changing consumer preferences in this respect, which play a crucial role in shaping demand-driven cross-price dependencies and, consequently, changes in spillovers that shape the transmission of shocks to individual prices to the CPI inflation rate.

\section{Conclusions}   \label{sec:conclusion}

The article introduces a novel Markov-switching spatial autoregressive (SAR) model with an unknown time-varying weight matrix and presents a Bayesian approach to statistical inference. The model aims to study the time-varying shocks' direct and indirect propagation in a system of interconnected variables (nodes). Using an unknown weight matrix allows more flexibility in capturing the connectivity patterns, while the temporal variation accounts for the cyclical behaviour and abrupt structural changes in the data. A Markov switching process drives the time variation of two main components of the SAR model: the elements of the weight matrix and, hence, the connectivity pattern among the response variables, and the spatial autocorrelation, which governs the speed of decay of a shock propagating through the system.

We illustrate the model's performance in an empirical application that examines consumer preferences' role in shaping demand-driven cross-price dependencies. The analysis identifies distinct states corresponding to significant economic events and shifts in consumer behaviour, with marked cross-country variation in the spillover from one price category to another. Overall, the proposed model provides a valuable framework for studying spillovers and understanding the links between economic variables and their variation over time.

There are various possible extensions to our proposed setup. For instance, the increasing use of stochastic-block models motivates a refinement of the time-varying element to particular blocks within the weight matrix \citep{hollandetal1983}. Related to this, an extension could then also be considered for community detection \citep{FORTUNATO201075}. Another possible direction for future research is to investigate different types of nonlinear spillover effects in the system, such as those incorporating threshold effects. Overall, these potential extensions and future research directions can further enhance our understanding of the spatiotemporal propagation of shocks and the dynamics of spatial dependence in interconnected economic systems.

Another interesting avenue for future research would be to extend the proposed Markov switching approach to the dynamic spatial autoregressive framework \citep[e.g., see][]{baltagi2014estimating,yang2021estimation,bille2022effect} to account for potential breaks in both the temporal and spatial dependence structures and strengths.
% Heterogeneous spatial weights \citep{aquaro2021estimation}.
% Grouped SAR panel model \citep{huang2023grouped}

%%%%%%%%%%%%%%%%%%%%%%%%%%%%%%%%%%%%%%%%%%
\bibliographystyle{chicago}
\bibliography{biblio.bib}

\end{document}